\documentclass[aps,prb,twocolumn,groupedaddress]{revtex4}
\usepackage{graphicx}
\usepackage{xcolor}
\usepackage{float}

\begin{document}

\title{Structures of solid hydrogen at 300K}

\author{Graeme J. Ackland$^1$ and  John S. Loveday}
\email[]{g.j.ackland@ed.ac.uk }

\affiliation{ $^1$Centre for Science at Extreme Conditions and School of Physics and Astronomy, University of Edinburgh, Edinburgh, U.K.}

\begin{abstract}
We present results predicting experimentally measurable structural quantities from molecular dynamics studies of hydrogen.  In doing this, we propose a paradigm shift for experimentalists -- that the predictions
from such calculations should be seen as the most likely hypotheses.
Specifically, the experimental results should be aiming to distinguish
between the candidate low-energy structures, rather than aiming to solve the simplest structure consistent with the data.  We show that the room temperature X-ray diffraction patterns for hydrogen phases I, III, IV and V are very similar, with only small peaks denoting symmetry-breaking from the hcp Phase I.  Because they incorporate atomic displacements the XRD patterns implied by molecular dynamics calculations are very different from those arising from the static minimum enthalpy structures found by structure searching.  Simulations also show that within Phase I the molecular becomes increasingly confined to the basal plane and suggest the possibility of an unusual critical point terminating the Phase I-III boundary line. 
\end{abstract}
\date{\today}

\maketitle

Solid hydrogen has proved to be one of the most challenging topics in high-pressure physics, both theoretically and experimentally.  At room
temperature, information about the crystal structure is available largely
through indirect methods such as
spectroscopy\cite{goncharov1998new,goncharov2001spectroscopic,loubeyre2002optical,dalladay2016evidence,howie2012mixed,zha2012synchrotron,howie2015raman,gregoryanz2003raman}. With the exception of two neutron diffraction studies \cite{somenkov,goncharenko2005neutron} at $\sim$ 30GPa, structural studies are confined to X-ray studies \cite{hemley1990equation,loubeyre1996x,akahama2010evidence,akahama2010raman}
which are largely insensitive to molecular orientation.  To 
exploit these studies fully, it is important to have models for the crystal
structure.  In recent years, {\it ab-initio} structure-search methods have
been highly successful at determining the possible classical ground
state structures\cite{pickard2007structure,pickard2009structures,pickard2012density,li2013classical,monserrat2016hexagonal}. 
These have shown a panoply of possible
phases, typically with large unit cells and low symmetry, often very
close in energy.  

The calculations have an unquantifiable uncertainty associated with the choice of functional\cite{azadi2013fate,clay2014benchmarking,drummond2015quantum,azadi2017role}.
Furthermore, the effects of quantum nuclear-motion are significant,
with zero-point energy being much larger than typical energy
differences between structures.  So despite all this work, no
consensus has emerged for the crystal structure of any high pressure
phase.  Nevertheless, some patterns have emerged which suggest the
calculated structures are consistent with the major experimental
findings\cite{magdau2017simple}.

Since the discovery of a Raman-active phonon, Phase I of hydrogen has
been accepted as a hexagonal close-packed (hcp) structure of rotating molecules. On cooling at
pressure, a transformation occurs to a ``broken symmetry'' Phase II,
characterised by a discontinuous change in H$_2$ vibron frequency and
the appearance of several low-frequency modes.  This transition occurs at temperatures
and densities where quadrupole interactions become significant, and
these are likely to be the driving force.  The I-II transformation has
no distinctive signature in X-ray diffraction \cite{loubeyre1996x}, suggesting that it is an orientational ordering of the hydrogen molecules on the hcp lattice. Many of the most stable candidate structures from density functional theory (DFT) calculation 
are in this category.\cite{pickard2007structure,geng2012high}

At higher pressures, above 160GPa at low-T,  pronounced weakening in
vibron frequency and further changes in the low frequency spectra
heralds Phase III. It is debatable whether there is any signature of
this phase in X-ray diffraction: at most it is a small distortion or
modulation of the hcp structure.  Perhaps the most distinctive
signature of Phase III is the sudden appearance of a strong infrared
signal, indicating that the structure has broken inversion
symmetry. At still higher pressures, darkening of the samples suggests
a bandgap closure in a molecular phase, and reflectivity reveals a 
transition to a metallic phase, predicted by DFT to be initially
molecular then a low-coordinated atomic solid.  Ultimately, hydrogen will
metallize and the molecular bonds will break, though it is unclear
whether these processes are simultaneous\cite{holst2008thermophysical,lorenzen2010first}.

At room temperature, the Phase I transforms first to phase III at around 180GPa, then to a Phase IV at around 230GPa which has not been observed at low temperature.  Phase IV is characterized by the appearance of a second, (and possibly
third) high frequency vibron\cite{howie2012proton,howie2012mixed,zha2013high,loubeyre2013hydrogen}.  Under further pressurization, the two
vibrons remain and changes in the low frequency Raman spectra may
indicate transformations to further phases IV' and V.  It is assumed
that metallization will occur, but this has not been observed at room
temperature.

We have spent several years making comparisons between DFT data and
the spectroscopic data, using lattice dynamics and molecular dynamics,
including path integral
methods\cite{magdau2013identification,magdau2014high,ackland2015appraisal,howie2014phonon,ackland2014efficacious,magdau2017simple,magduau2017theory,magdau2017infrared}.  We have used different exchange-correlation functionals and
different treatments of anharmonicity, and our conclusion is that
these methods are not sufficiently accurate to obtain quantitative
agreement for transition pressures or vibrational frequencies.  Nevertheless, in this paper we will present some
 predictions about crystal structures which are experimentally measurable.

\subsection{ab initio molecular dynamics}

Structure search algorithms work well for low temperature phases with
harmonic phonons, but even at room temperature hydrogen is far from
harmonic.  Ab initio molecular dynamics (AIMD) is able to probe this region.
Limitations on timescales and system sizes mean that accurate free
energy calculations are impossible, however, just as with structure
search, AIMD can reveal symmetry and structure of {\it candidate} phases.

Most previous AIMD was done with the PBE functional \cite{PBE}.  However, it is now becoming obvious that this {\it de facto} standard functional has a specific failing: it
overstabilizes metallic structures relative to molecular ones. This can
be traced to a design feature - PBE does not reproduce the energy in
the limit of large $\nabla\ln(\rho)$.  This does not usually cause
problems: when studying metallic phases, the high $\nabla\ln(\rho)$
regime is not sampled, and in comparing molecular phases the error
cancels out.  It is only in the specific case of a molecule-metal
transition that it becomes critical.  In this work we also use the BLYP
functional\cite{blyp-LYP,blyp-B} which, though simple, does capture the high
$\nabla\ln(\rho)$ limit and gives a better description of H$_2$ 
molecular dissociation. 

\subsubsection{Phases}
Experimentally, four numbered phases have been
reported based on spectroscopy.  In addition, two ``primed'' sub-phases
have been identified, giving a sequence I-I'-III-IV-IV'-V.  The broken
symmetry phase II and metallic phases have been observed only at at low
temperatures.

Previous MD on phases of hydrogen at 300K suggests only Phase III
involves harmonic (or even anharmonic) oscillations about well-defined
atomic positions: all other phases have molecular rotation, reorientation
and at higher pressures significant rebonding.  All of them can be
characterised by molecular ``motifs'' located on ``hexagonal close
packed'' lattice sites.  This underlying P6$_3$/mmc space group
provides the highest possible symmetry - changes to the motifs lower
this symmetry. 

Surprisingly, previous calculations were done in the NVT ensemble,
so that the crystallographic measurable, the $c/a$ ratio, has not
previously been calculated.  For close-packing of hard spheres, the
c/a ratio is $\sqrt{8/3}=1.633$.  Cohesion in solid hydrogen arises
primarily from van der Waals forces which drop off as $1/r^6$.  The
Lennard-Jones potential captures this behavior, and stabilizes the 
hcp structure with  $\sqrt{8/3} \simeq 1.633$.\cite{loach2017stacking}

\subsection{AIMD runs}
\begin{figure}
\includegraphics[width=100mm]{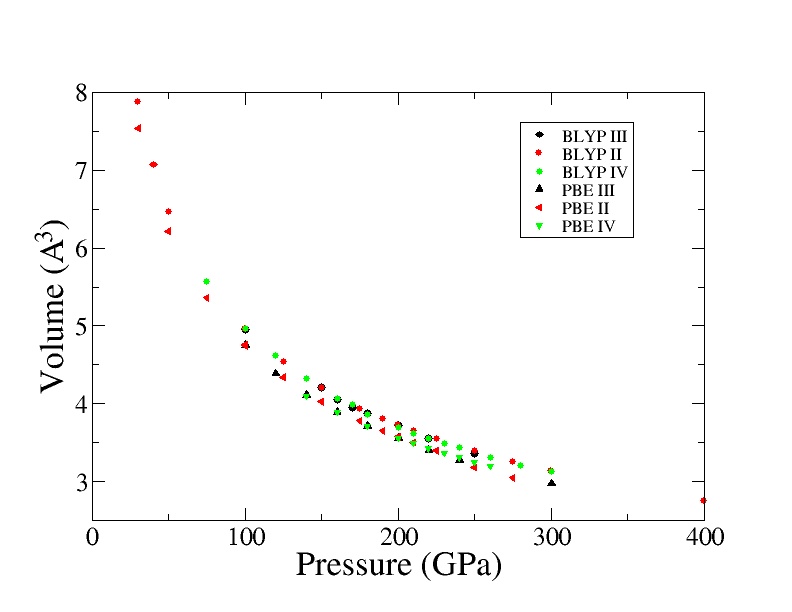}
\caption{Equation of state (volume per atom) for all structures with BLYP (circles) and
  PBE (triangles) showing that functional effects are much larger than
  structural differences, and the uncertainty due to functional is about 20GPa.  }
\label{fig:EoS}
\end{figure}

We ran a large number of molecular dynamics calculations to evaluate
the various structures. The  same sequence of phases are observed independent of exchange-correlation functional. Compared with PBE, the BLYP functional gives 
systematically higher
pressures at a given density (Fig.\ref{fig:EoS}).  It also makes better defined hydrogen molecules with higher vibrational frequencies.

 Calculations were initiated from different
candidate structures identified from previous {\it Ab-Initio} Random Structure Search, AIRSS, calculations for phase
II, III and IV candidates\cite{pickard2007structure,pickard2009structures}.  None of those low-symmetry structures
remained stable at 300K, all transformed to higher symmetry
structures.  Nevertheless, based on average molecular positions 
some distinct structures were observed
which can be assigned to non-metallic phases I, III, IV, V plus a molecular 
metallic phase $Cmca$ and atomic metal $I4amd$.

\subsubsection{Finite Size Effects}

The complexity of phases III, IV and V mean that they are extremely
sensitive to finite size effects. As shown in Table \ref{table1}, only
phases compatible with the initial conditions are observed.  e.g. the
$BG'BG''$ phase IV candidate is hexagonal a four layer repeat with six
atoms per layer - self evidently, only super-cells with multiples of 24
atoms can find this structure, while the $BG'_xBG'_yBG'_z$ candidate 
requires a multiple of six hexagonal layers. 

Furthermore, there is a probability of finding a layer with incorrect
stacking.  This is of order $\exp(-N\Delta F/kT)$, where $\Delta E$ is
the excess free energy per atom in the mis-stacked layer and N the
number of atoms per layer.  Evidently, this goes to zero at large $N$
- {\it mis-stackings never occur in thermodynamic equilibrium}.
However, in finite systems it may happen: with 12 atoms per layer,
even fluctuations between $B$ and $G$ occur. We found that with less
than 54 atoms {\it per layer} spurious fluctuations between types of
$G$ layer at the size of the system do still occur, which gives a
spuriously high mean-squared displacement.  

For a simulation to even have the {\it possibility} of correctly
describing Phase IV, it should accommodate both $BG'BG''$ and
$BG'_xBG'_yBG'_z$ candidates, and have layers containing a multiple of
6 atoms.  To also prevent spurious fluctuations required a minimum of
648 atom (i.e. 54 atoms per layer).  This cell size was used in the
region of the phase transition.

Finite size effects are generally regarded as a problem, but if
properly understood they can be turned to advantage.  Specifically, by
adjusting the cell size to be incompatible with the stable phase, we
can probe metastable phases.  This enables us to predict experimental
signatures for all candidate phases, and thus determine whether they
could be distinguished by diffraction or spectroscopy. 

All the cells considered can transform into Phase I,
which allows us to determine whether the transitions are first order.
We monitor three order parameters, the density, the $c/a$ ratio and
the angle between the molecules and the c-axis.

\begin{figure}
\centering
\includegraphics[width=1\columnwidth]{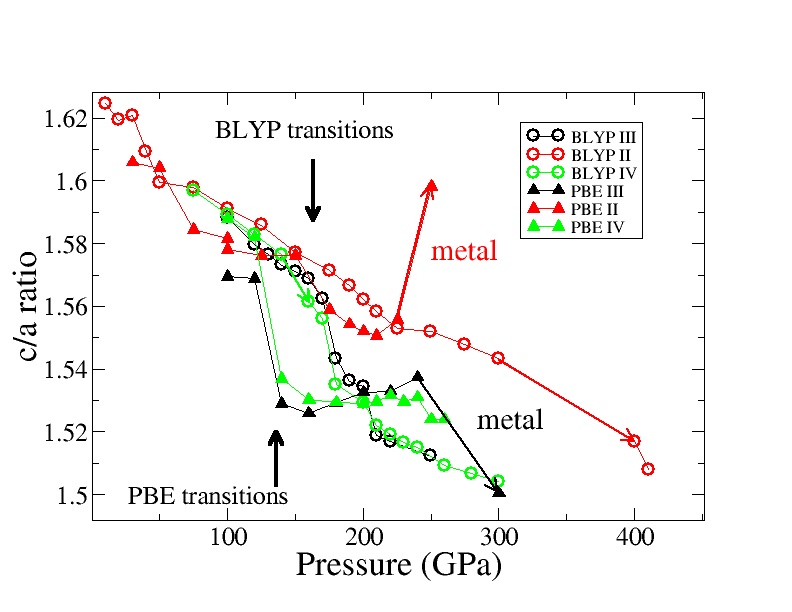}
\caption{Plot of c/a ratio for simulations in various cells.  Arrows
  show sharp change in c/a with transition to phase IV, and more
  gradual change approaching phase III.  Note the significant
  functional dependence in the calculated transition pressure.  The
  very high pressure metallic C$mca$ structures are twinned, so the change in "$c/a$
  ratio" for the cell signifies the transition, but is not the $c/a$ ratio of C$mca$. }

\label{covera}
\end{figure}

\begin{figure}
\includegraphics[width=100mm]{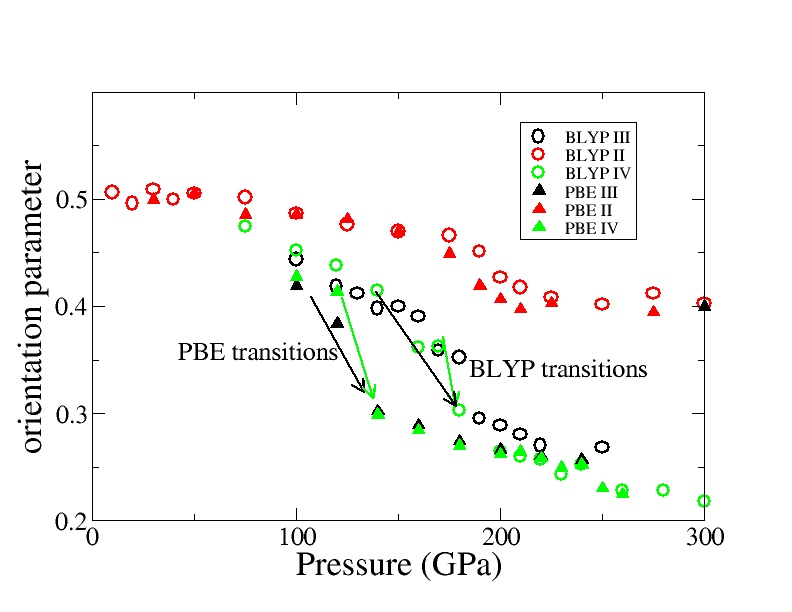}
\caption{
Variation of angular orientation order parameter $<\cos\theta>$ with pressure.
}
\label{fig:cos}
\end{figure}

At the lowest pressures Phase I comprises freely rotating H$_2$ molecules.
Fig.\ref{covera} shows that, as expected, at low pressures $c/a$ tends
to the ideal ratio but gets smaller under pressure. To understand why this might be, we examined the
cosine of $\theta$ between the molecular axis and the $c$-axis
(Fig.\ref{fig:cos}).  For a free rotor, this would average 0.5.  This
is the case at low pressure, but even within Phase I, as the pressure
increases, the molecule increasingly rotates in the plane.  This
reduction of $c/a$ has been observed by X-ray differaction\cite{loubeyre1996x,akahama2010evidence} at low temperature,  and can
now confidently be ascribed to the molecule changing from spherical to
toroidal.  The torus is still compatible with the P6$_3/mmc$, so this
symmetry breaking of the molecule does not require a structural phase
transition.

This change from spherical to torus rotation is not observed in NVT
ensemble simulations with ideal $c/a$, emphasizing the importance of
choice of ensemble.

\begin{figure}
\centering
\includegraphics[width=0.48\columnwidth]{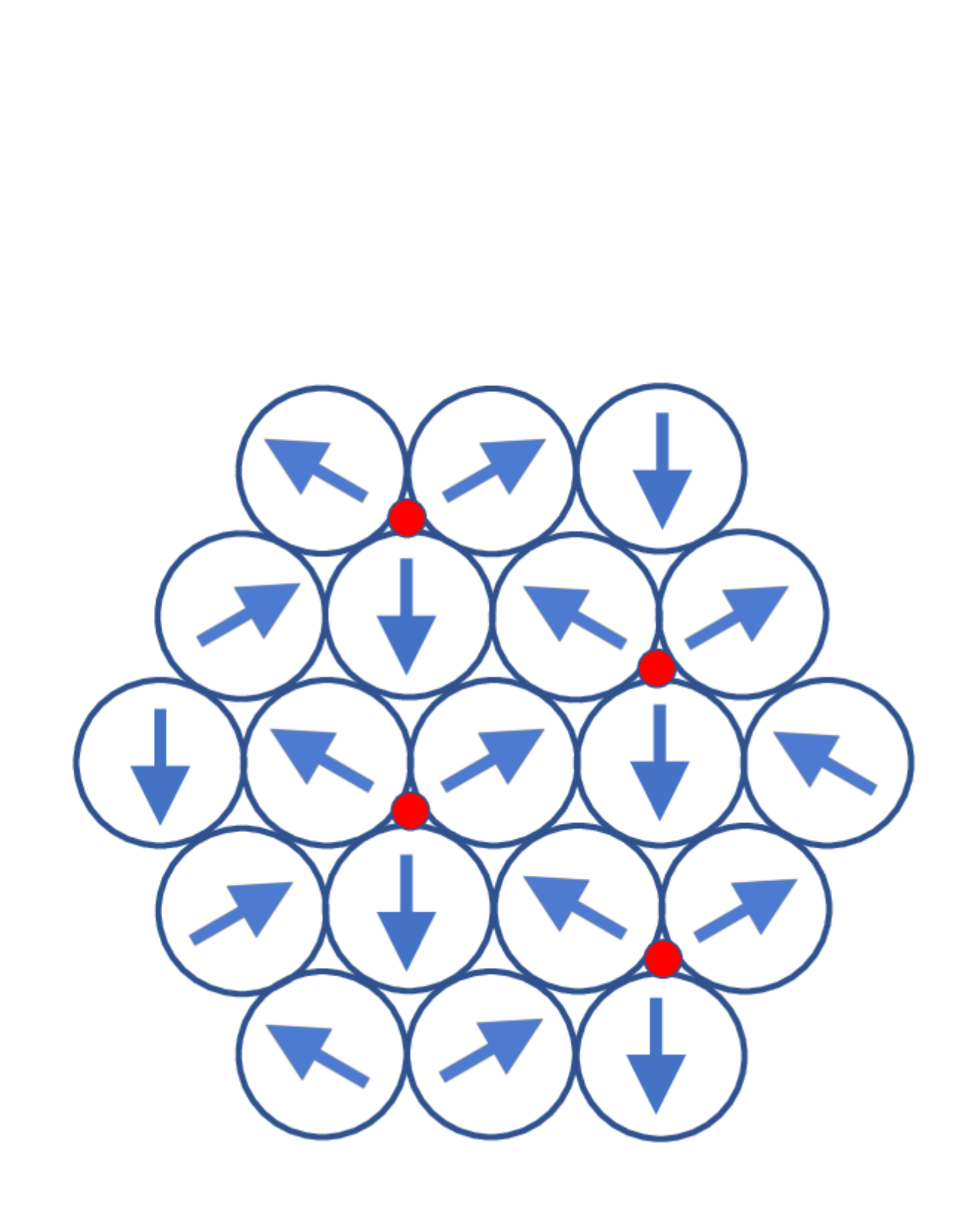}
\includegraphics[width=0.48\columnwidth]{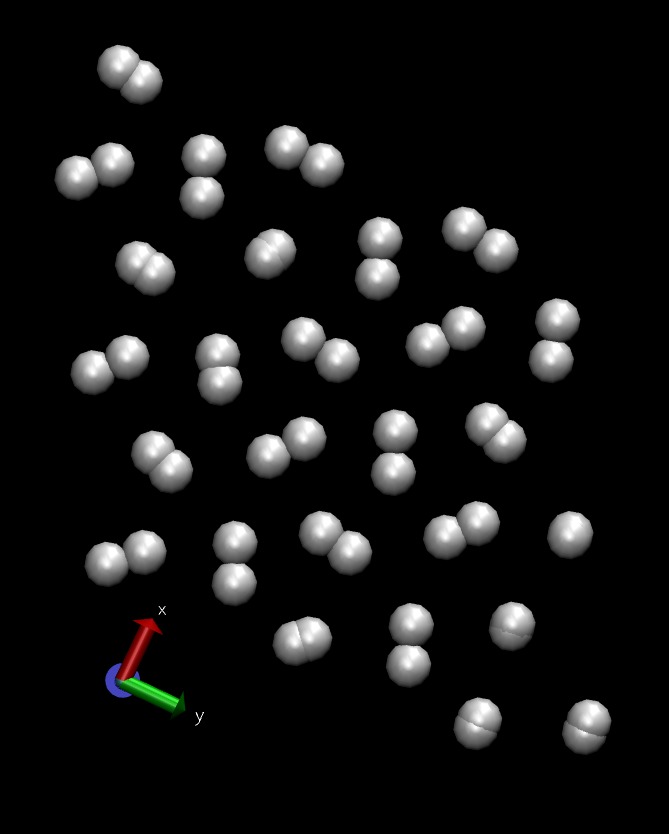}
\caption{Single $G$ layer, All structures comprising stacking of these
  layers are labelled ``Phase III'.  (left) schematic: red dots show
  position of the 3-fold rotation axis.  Blue arrows represent
  molecular axis. The orientation within the plane does not require
  any symmetry-breaking, however any orientation of molecules in
  planes above and below will break the symmetry and induce a dipole,
  as indicated by the direction the arrow.  (right) time-averaged positions
  of atoms aver 1ps from one layer in the 180GPa simulation using
  BLYP.  All molecules are in-plane, apparently short bonds occur when
  the molecule has rotated through 180 degrees at some stage.  Notice
  how molecules are displaced.}
\label{PhaseIII}
\end{figure}

At higher pressures there is a transformation to Phase III.  Structure
searching has revealed a number of candidate structures which were
initially described by reference to the nuclear positions as different
stackings of ``distorted Graphite-like layers''.  However, considering the
molecular (rather than atomic) positions reveals that this is just an hcp lattice with the minimum of broken
symmetry required for molecular orientation (Fig.\ref{PhaseIII}(a).  The molecular dynamics shows a similar orientational order (see Fig.\ref{PhaseIII}b).

\begin{figure}
\centering
\includegraphics[width=1\columnwidth]{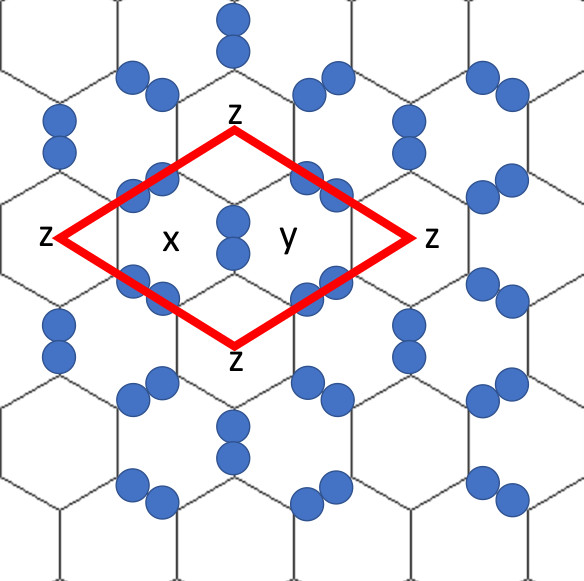}
\includegraphics[width=1\columnwidth]{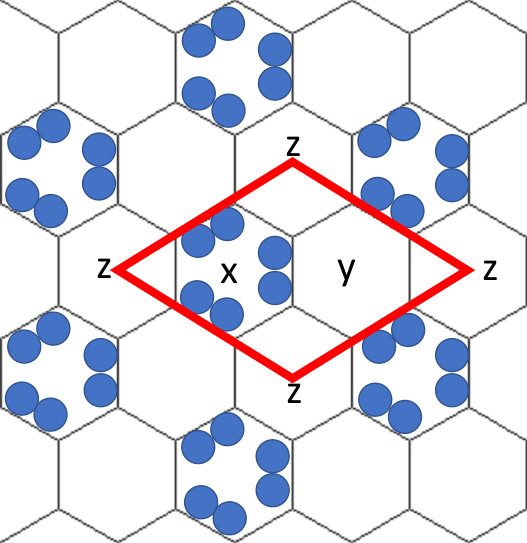}
\caption{Schematic of typical in-plane atomic positions in the so-called
  ``graphitic'' ($G$) layers of Phase IV. The red diamond shows $a$ and $b$
  vectors for single unit cell. Each unit cell has two equivalent and
  one non-equivalent hexagon: $x$, $y$ and $z$ are used to label the
  layer stacking of the non-equivalent site. In this notation (a)
  $G'_z$, (b) $G''_x$.  The molecular ``B''-layers are like Phase-I
  and are not shown, they simply comprise a molecule at the center of
  each hexagon, again six atoms per layer. In static relaxation these
  B-molecules have well defined orientation (e.g. the Pbcn
  structure), but at room temperature they are disordered and
  re-orient on a 100fs timescale.  Due to constraints from periodic
  boundary conditions, in MD simulation a two-layer cell in Phase IV
  PT conditions adopts $BG'_z$ stacking, four layers $BG'_zBG''_z$,
  six layers $BG'_xBG'_yBG'_z$, eight layers $BG'_zBG''_zBG'_zBG''_z$.
\label{fig:BG}}
\end{figure}

To understand the highest pressure structures, fig. \ref{fig:BG}
relate the observed structures to the simple MgB$_2$ structure with a
hydrogen molecule on the Mg site (a triangular ``B'' layer) and
hydrogen atoms on the boron sites ( a graphitic layer ``G'').  This
structure has alternating layers, so the c-glide symmetry is broken
and the space group becomes P$6/mmm$.  In the molecular dynamics, this
MgB$_2$ structure is recognised {\it on average} at very high
pressures.  However it is energetically highly unstable to formation
of molecules: the trajectories cannot be described in terms of
harmonic oscillations: the boron lattice site is not a local maximum
of energy.

The structures observed for phases IV, IV' and V are described in terms of
symmetry-breaking from MgB$_2$ so as to form molecules in the layers. 
There are multiple ways of doing this (Figure \ref{fig:BG}).  
The molecules in these layers
tend to remain in plane, meaning that the c/a ratio falls further.

In figure \ref{PhaseIII} we show a schematic of how a layer in Phase
III emerges from Phase I.  The large circles represent molecular
locations on a perfect close packed plane.  We observe that the
molecules in Phase I come to lie in the plane at high pressure.  Now,
assume that each in-plane molecule points towards a gap between
neighbour and is not aligned with its neighbours.  These two rules are
sufficient to uniquely define all the molecule orientations, as shown
by the arrows.  Figure \ref{PhaseIII} also shows a picosecond
time-average from 648-atom BLYP simulation at 180GPa, assigned Phase
III.  Although the non-centrosymmetric motif is clear, there are
frequent local rotations and reorientations.

This ordering leads to a 3-molecule repeating cell, and spontaneously
breaks inversion symmetry.  This broken symmetry means that the
molecule moves off the hcp site and acquires a dipole moment. This
dipole moment is responsible for the strong IR signal which is a key
signature of Phase III.  The movement off-site might be detectable by
X-ray scattering, but new peaks associated with it are weak, and it
only induces a small change in relative intensity of the three main
peaks compared with hcp.

Furthermore, there are two non-equivalent sites for the next layer
(2/3rds unmarked, 1/3rd red circles).  Consequently, a 3D unit cell
must be based on (at least) two of these 3-molecule 2D layered cell
The lowest energy structures identified by ab initio structure search
for Phase III, C$2/c-24$ and P6$_1$22, involve a 4 and 6 layer repeat of
this layer.

Determination of the high temperature structures was done by layer-by
layer {\it ab oculo} analysis using vmd.  In addition to snapshots or
movies proved, two analyses proved  extremely useful.
\begin{itemize}
\item Plots of time-averaged atomic positions.
  The $B$-layers image as a triangular lattice with two atoms
  coincident at each lattice site.  The $G''$ layers image as a large
  triangular lattice with six atoms coincident at each site.  The $G'$
  layers typically image as separate atoms, similar to a snapshot,
  although after many picosecond the pattern is destroyed by diffusion within
  this layer.

\item Dot-plots for all atoms, at all times.  The $B$-layers image as
  spheres or small donuts, the $G''$ layers image as triple-arcs or
  large donuts, with some evidence of six- and three-fold rotational
  symmetry, the $G'$ layers image as separate atoms.

\end{itemize}

We carried out limited path Integral molecular dynamics, which 
show relatively little qualitative
change from the classical picture, the main effect being a wider
variation in molecular length due to zero-point energy.  There is some
small effect on the phase boundaries.

\subsubsection{Simulated Crystallography}

We have calculated the diffraction pattern from the positions of the atoms from a sample of MD runs. in phases I, III, IV and V.  This was done by combining the positions from the final 2ps of the MD trajectory and treating the supercell as a single cell with P1 symmetry. The resulting assemblage of approximately 50000 atomic positions thus models not only the average positions of the atoms within the structure but also the atomic displacements, including anisotropy and anharmonicity, about these average positions. The calculations were done using the GSAS-II program \cite{gsasii} and assumed a standard hydrogen form factor and  an X-ray wavelength of 0.7 {\AA}.

\begin{figure}
\includegraphics[width=\columnwidth]{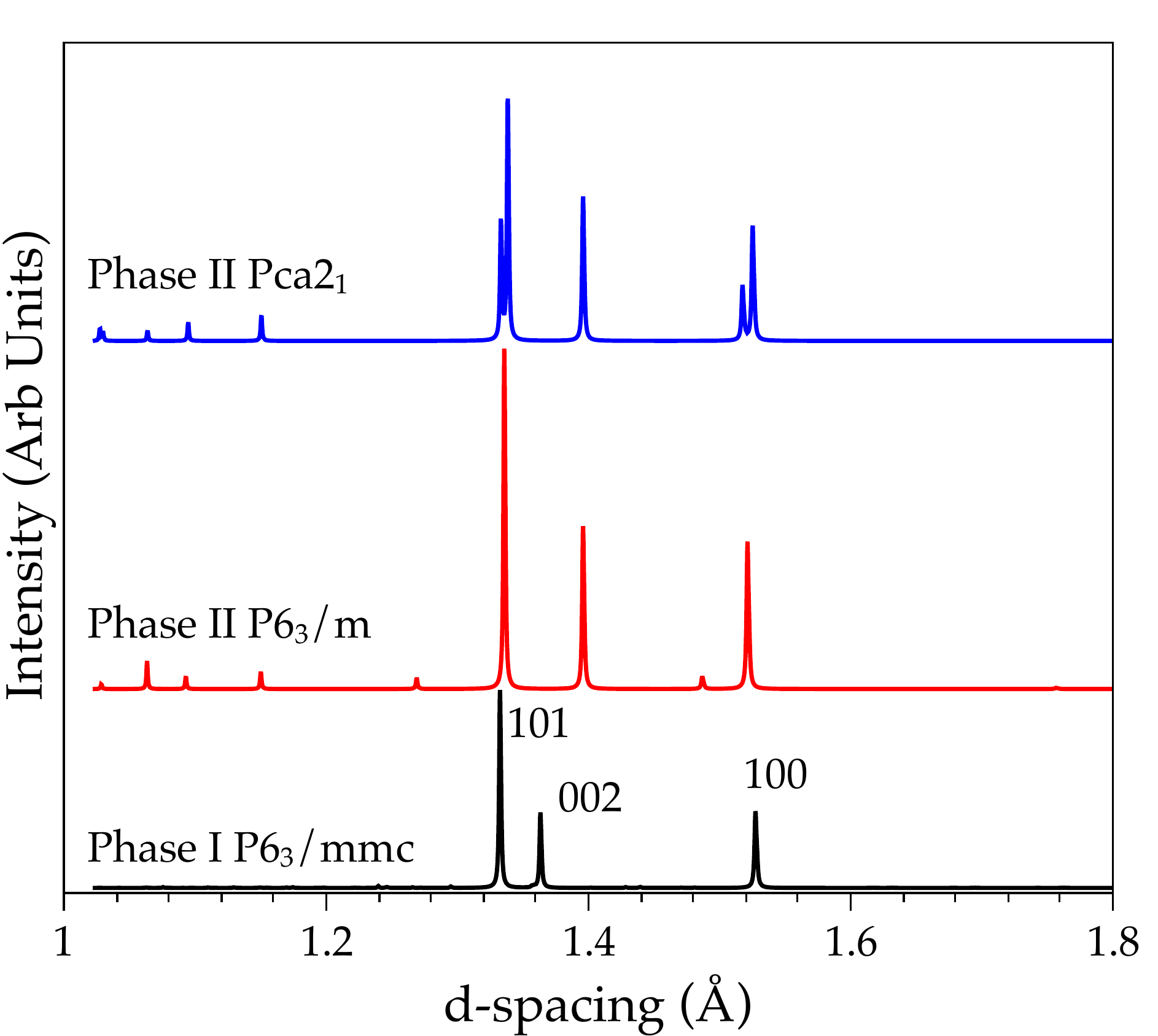}
\caption{
Simulated X-ray diffraction patterns for Hydrogen Phase II at 140GPa.  P$6_3/m$ and P$ca2_1$ are the zero-temperature ground states proposed by structure search.  MD averaged over 2ps starting from P$6_3/m$ at 300K temperature is shown.  Simulations starting from  P$ca2_1$ or P$6_3/mmc$ are indistinguishable - all are Phase I.
}
\label{fig:xrdii}
\end{figure}

Figure \ref{fig:xrdii} shows that at 140GPa, the XRD patterns for two lowest energy candidate structures for Phase II P6$_3$/m and Pca2$_{1}$ are similar, the distinguishing feature being a small peak splitting in Pca2$_{1}$ or small additional peaks in P6$_3$/m.   XRD of either supercell gives 3 significant diffraction peaks, which can readily be indexed as (100) (002) and 101) from an hcp structure.  In the 300K MD, simulations started in either structure transform to an identical hindered-rotor hcp Phase I.

\begin{figure}
\includegraphics[width=\columnwidth]{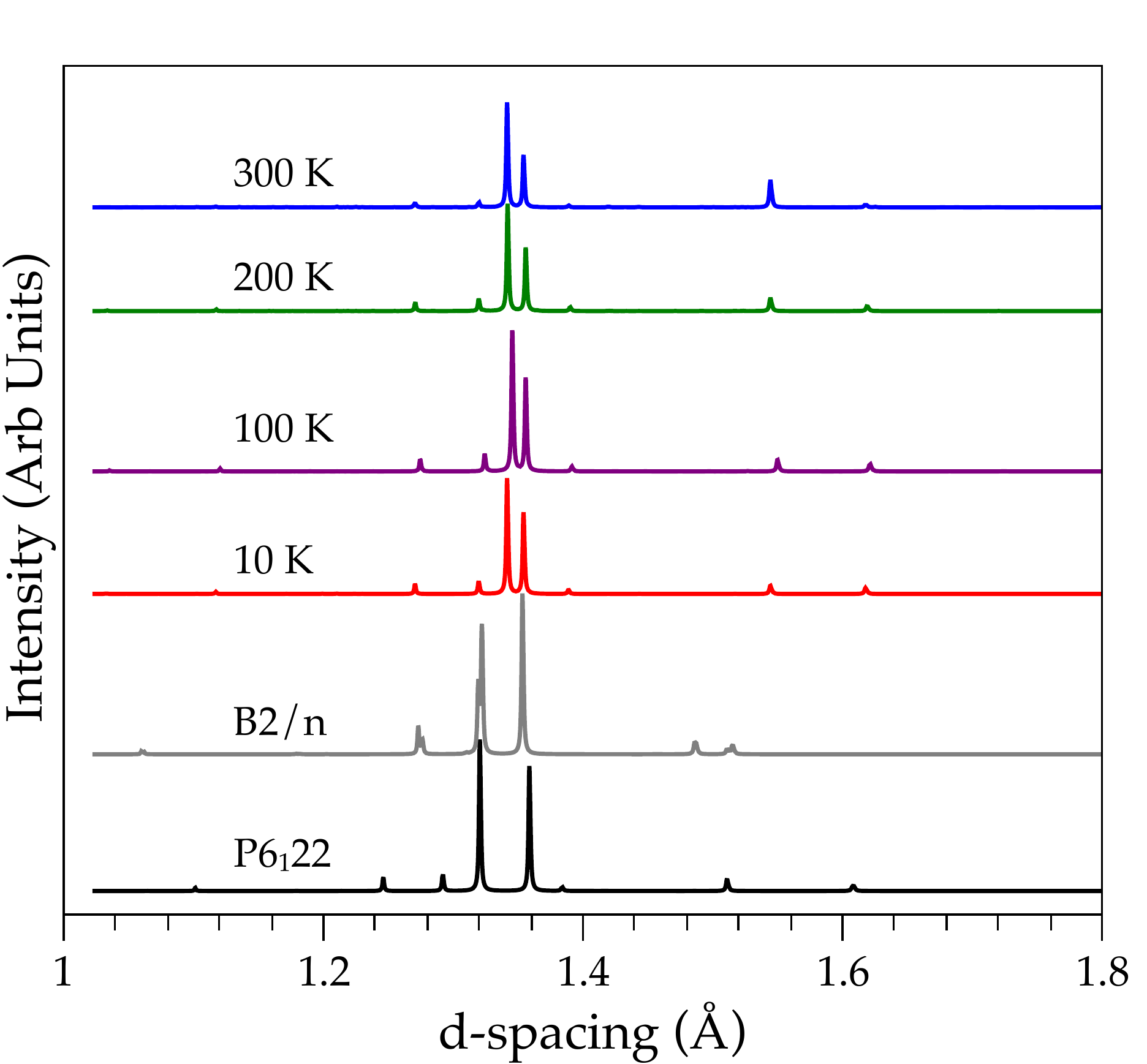}
\caption{
Simulated X-ray diffraction patterns for Hydrogen Phase III at 190GPa calculated using BLYP.  B$2/n$ and P$6_122$ are the zero-temperature ground states proposed by Pickard\cite{pickard2007structure} and by Monserrat\cite{monserrat2016hexagonal} respectively.  MD is averaged over 2ps starting from P$6_122$ at various temperatures.
}
\label{fig:xrdiii}
\end{figure}

Phase III is stable at 190GPa, and Fig.\ref{fig:xrdiii} shows that again the two zero-temperature candidate structures P$6_122$ and B$2/n$ (sometimes called by its alternative setting C$2/c$) have similar 2-peak patterns, being distinguished only by weak reflections.  In the MD, the two-peak pattern persists, but unusually as the temperature increases, a third small peak grows in prominence, while other small peaks vanish.   On simulated cooling, the XRD pattern transforms continuously from being characteristic of phase I, to characteristic of Phase III, which is consistent with the gradual onset of short ranged orientational order.

\begin{figure}
\includegraphics[width=\columnwidth]{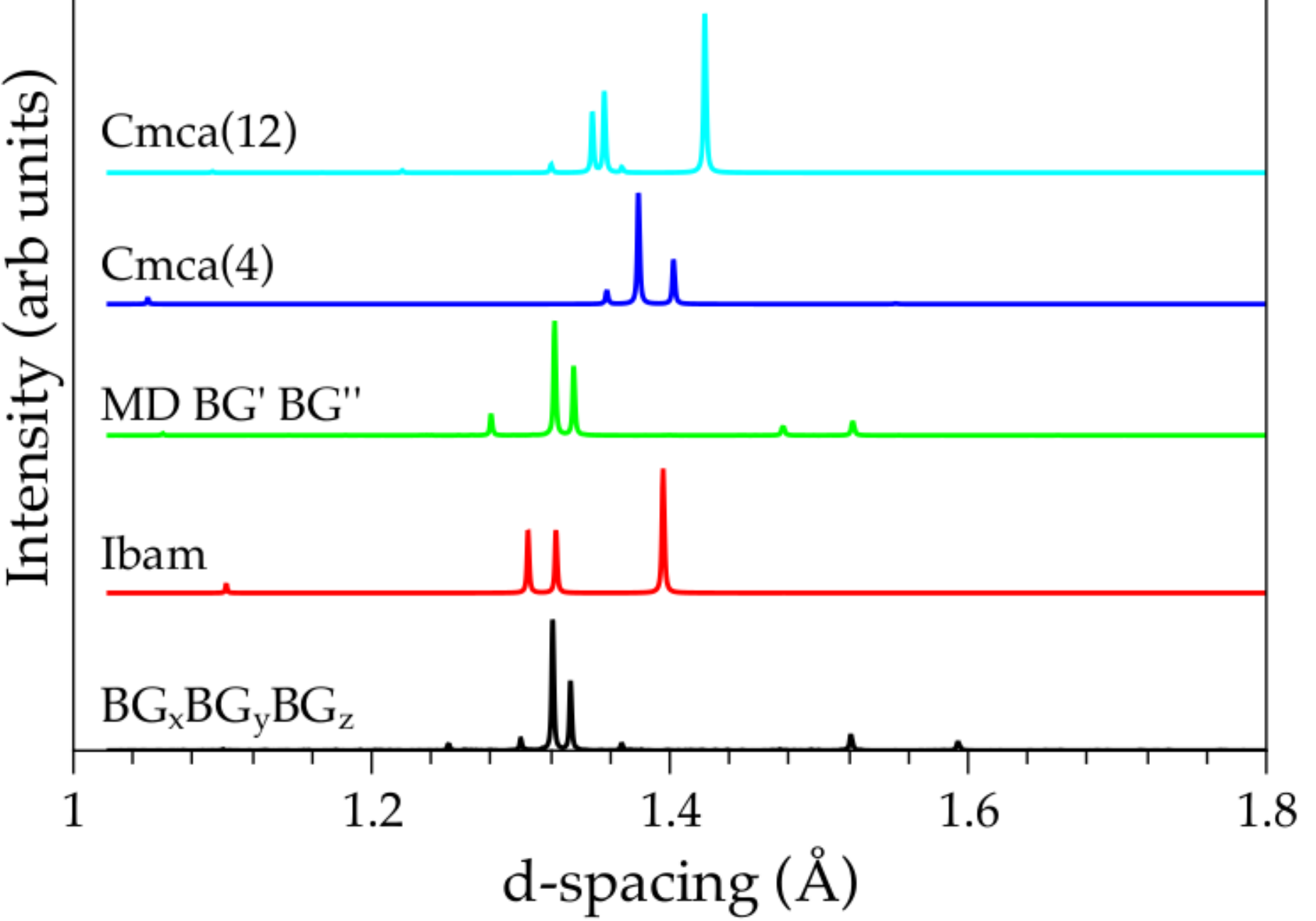}
\caption{
Simulated X-ray diffraction patterns for Hydrogen Phase IV at 220GPa. Ibam Pc and Pbcn are the zero-temperature ground states proposed by Pickard.  MD for the two different hexagonal candidates (Fig. \ref{fig:BG}),  is averaged over 2ps at 300K starting from P$bcn$  ($BG'BG"$) and starting from  P$6_122$ ($BG_xBG_yBG_z$).
}
\label{fig:xrdiv}
\end{figure}

There are several candidates for Phase IV, which can be simulated separately by making supercells incompatible with the other.  Fig.\ref{fig:xrdiv} shows the calculated diffraction pattern from two of them, and again the two candidate give near identical patterns with a very close doublet and a third peak at lower angle. Ignoring the small peaks, would be possible to index these peaks to hcp, with an anomalously small $c/a$ ratio as shown in Fig.\ref{covera}.   Fig.\ref{fig:xrdiv} also shows that a very strong additional peak appears with the Ibam structure which corresponds to symmetric graphitic layers, a candidate for Phase V which is stable in MD above 400GPa\cite{magdau2013identification,magdau2016phaseV}.  This is perhaps the first phase which can be easily distinguished from hcp with XRD data.

The original 1935 prediction of metallic hydrogen by Wigner and Huntingdon\cite{wigner1935possibility} is based on free electron theory, and analysis of metallic hydrogen is still based on this premise\cite{dias2016new}. Since hydrogen has no core electrons, a
free-electron phase of solid hydrogen would have a featureless X-ray
diffraction pattern. However, calculations using DFT show that the
electrons are still well localized, and X-ray diffraction from metallic
hydrogen will be nearly as strong as from molecular phases\footnote{Ingo Loa, private communication}.

Very recently, a report of the experiments simulated in this work has appeared\cite{ji2019ultrahigh}.  In this paper, the sharp turn of the c/a ratio at the transition from Phase I - IV at around 200GPa is observed.  This is the signature of the phase transformation from hcp to a multilayer hexagonal structure, with the appearance of a second vibron.   The experiment rules out any of the lower symmetry zero-temperature structures found from structure searching as being candidates for Phase IV.

 Ji {\it et al}  attribute their results to a P$6_3/mmc$ structure. This would be a unique example of an isostructural electonic transition between two non-metallic phases.
 We have reproduced the band structure and recalculated this structure using BLYP,  find it to be energetically unstable at all pressures, as it is with PBE.  The associated variation in $c/a$ is shown in Fig. \ref{fig:cova} and diffraction pattern in Fig.\ref{fig:xrdiii}, both are in qualitative disagreement with  Ji {\it et al}'s own XRD data: in particular the c/a ratio of the proposed structure {\it increases} with pressure, whereas the XRD indicates a decrease.  Furthermore, the P$6_3/mmc$ structure in unstable in MD, transforming spontaneously to the normal Phase I, or lower symmetry III and IV structures previously identified.

\subsection{Discussion and Conclusions}

We have carried out extensive molecular dynamics simulations of high pressure hydrogen at room temperature using two different exchange correlation functionals.  The functionals give the same sequence of phases, but with a difference of 20-30GPa in pressure.   The sequence of phases, I-III-IV-V is in accordance with experiment, with the pressure calculated using in BLYP closer agreement.  Molecular rotation (or disorder) increases the symmetry so that the calculated diffraction pattern for Phases III and IV is much more similar to Phase I than to their zero-temperature relaxed structures previously taken as exemplars.

Under pressure the free rotors of Phase I become more and more inhibited, with the molecules preferentially rotating in-plane.  This loss of sphericity causes a drop in the c/a ratio away from ideal.  In phase III, the rotation stops and the molecules lie in plane, however the diffraction pattern structure is still close to hcp with a still-lower c/a ratio.  A 3\% drop in the c/a ratio and change in its pressure-slope accompanies the transformation.

Our simulated XRD patterns, and the implied variation in the c/a ratio, are in excellent agreement with recent XRD data.  The room temperature, high pressure phases of hydrogen can therefore confidently be ascribed to hexagonal structures with inhibited rotors, up to at least 250GPa.


GJA  acknowledges the support of the European Research
Council Grant Hecate Reference No. 695527

\vspace*{-5mm}
\bibliographystyle{apsrev} 
\bibliography{Refs}
\vspace*{-10mm}

\newpage
\appendix{Appendix}

\begin{table}
\centering
\begin{tabular}{|l|l|l|l|l|l|}
\hline
BLYP \\
\hline
Pressure & Volume & c/a & $<cos\theta>$ & initiated & Phase \\
\hline
10 & 11.9804 & 1.6247 &  0.506480 & P63m  & I \\
20  &9.1007  & 1.619415 & 0.4962698 & P63m  & I \\
30  &7.8787  &1.620801  & 0.508926 & P63m & I \\
40 & 7.0702 & 1.609334 & 0.4996069 & P63m & I \\
50  & 6.4600 & 1.599574  & 0.5051744 & P63m & I \\
75 & 5.5644 & 1.597964    & 0.5018897 & P63m & I \\
100  & 4.9660  & 1.591320  &0.4865163& P63m & I \\
125  & 4.5356  &   1.585926 & 0.4764345& P63m & I \\
150  & 4.2018  & 1.577247  &  0.4698209& P63m & I \\
175  & 3.9393  & 1.571384 & 0.4662826 & P63m & I \\
190 & 3.8078  & 1.566554   & 0.4507833& P63m & I \\
200 &  3.7320 &  1.562305 &  0.4266832 & P63m & I \\ 
210 & 3.6532  &  1.558341  & 0.4177813 & P63m & I \\
225 & 3.5491  & 1.553017  & 0.4081023 & P63m & I \\
250 &  3.3899  & 1.552009 & 0.4016243 & P63m & I \\
275 & 3.2574 & 1.547899  & 0.41143 & P63m &I \\
300 & 3.1358 & 1.54326   & 0.4024638 & P63m &I \\
400 & 2.7490 &  1.517 &  0.3613476   &  P63m & Cmca\\
410 & 2.6943 & 1.508 & 0.442654  &  P63m & Cmca\\
\hline
100 & 4.9604 & 1.5883 &  0.4439265 &  P6$_1$22 & I  \\
120 &4.6034 & 1.579752 & 0.4185463 &  P6$_1$22 & I  \\
130 & 4.4539 & 1.576561&  0.4116525  &  P6$_1$22 & I  \\
140  & 4.3176 &1.573193&  0.397990 & P6$_1$22 & I  \\
150 & 4.1895 & 1.571013 &  0.402121 &   P6$_1$22 & I \\
160 & 4.0992 & 1.568875  &  0.3899782 &  P6$_1$22 & I \\
170 & 3.9935 & 1.562581 & 0.3584516 &   P6$_1$22 & I \\
180 & 3.8947 & 1.556153  & 0.3523557 &   P6$_1$22 & III*  \\
190  & 3.7827 & 1.536208 &  0.295101  & P6$_1$22 &  IV \\
200 & 3.7258 & 1.534587 & 0.3058081  &   P6$_1$22 &  III  \\
210  &   3.6255 &    1.518815 &  0.2798709 & P6$_1$22& III    \\
220 & 3.5560 & 1.516970 & 0.2702824 &   P6$_1$22 &  III \\ 
250 & 3.3657 & 1.512323 &   0.267668 &    P6$_1$22 & III    \\
\hline
180a  & 3.8671 & 1.543891 &    0.3188191 & 648 P6$_1$22 & I    \\
180  & 3.8718 & 1.54544 &   & 648 P6$_1$22 & I    \\
200 & 3.3657 & 1.5024375  &   0.2883576 &   648 P6$_1$22 & GxGyGz    \\
220 &  &   &    &   648 P6$_1$22 & GxGyGz \\
\hline
075 &5.5675 &  1.596835 & 0.4742624 & Pbcn & I \\
100 & 4.9652 & 1.589387 & 0.4523625  & Pbcn & I \\
120 &  4.6145 & 1.582851 & 0.4385514   & Pbcn & I \\
140 & 4.3276 & 1.576478 & 0.418559 & Pbcn & I \\
160 & 4.0724 & 1.561678 & 0.3615409 & Pbcn & I  \\
170 & 3.9885 & 1.556033 & 0.3630052 & Pbcn & I\\
180 & 3.8619 & 1.535091 & 0.302474 & Pbcn & I\\
200 & 3.6941 & 1.529361 & 0.264194 & Pbcn & IV \\
210 & 3.6246 & 1.522132 & 0.2599856 &Pbcn & IV\\
220 & 3.5605 & 1.519139 & 0.257028 & Pbcn & IV  BG''BG''\\
230 & 3.4960 & 1.516730 & 0.2434004 & Pbcn & IV\\
240 & 3.4382 & 1.515042 & 0.2521973 & Pbcn & IV\\
260 & 3.3116 & 1.509366 & 0.2279203 & Pbcn & IV\\
280 & 3.2103 & 1.506743 & 0.2278928 & Pbcn & IV \\
300  &3.1251&  1.504065 & 0.2176326  & Pbcn & IV \\
\hline
 \end{tabular}
 \caption{Data for BLYP
 \label{table1}
 }
 \end{table}

\end{document}